\documentclass[fleqn,12pt]{scrartcl}
\usepackage{geometry}
\usepackage{epsfig}
\begin{document}                % INITIALIZE - DONT CHANGE
\title{ Programming matrix optics into Mathematica }
\author{Jos\'e B. Almeida\\
\emph{Universidade do Minho, Escola de Ci\^encias}\\ \emph{
4700-320 Braga, Portugal}}
%\date{}

% \author{}   % Use this and the next line only if there is a second
% \address{Another University, etc.}  % address. (Remove the left % marks)
%
\maketitle
\begin{abstract}
The various non-linear transformations incurred by the rays in an
optical system can be modelled by matrix products up to any desired
order of approximation. Mathematica software has been used to find
the appropriate matrix coefficients for the straight path
transformation and for the transformations induced by conical
surfaces, both direction change and position offset. The same
software package was programmed to model optical systems in
seventh-order. A Petzval lens was used to exemplify the modelling
power of the program.
\end{abstract}
\section*{Keywords} Aberration, matrix optics,
computer optics.
\section{Introduction}
In previous papers \cite{Almeida98, Almeida99} it was shown that it
is possible to determine coefficients for matrix modelling of
optical systems up to any desired order, computing power being the
only limiting factor. The second of those paper lists the calculated
seventh-order coefficients for systems comprising only spherical
surfaces.

The use of matrices for optical modelling of optical systems under
paraxial approximation is well known, see for instance
\cite{Gerrard94}. Other authors have shown the possibility of
extending matrix modelling to higher orders \cite{Kondo96, Laksh97}
and the author has used in his previous work a different set of
coordinates which made feasible the implementation of matrix models
in personal computers.

The first part of the paper describes the calculations necessary for
the determination of matrix coefficients and the manner these were
performed with Mathematica software \cite{Mathematica}. The
programming is general for surfaces that can be described
analytically and for whatever approximation is desired. The
possibilities of the modelling system are then exemplified with the
study of a Petzval lens' aberrations.

A set of Mathematica functions needed for the implementation of
matrix models was defined and assembled in a package which is
described in its most relevant aspects in the appendix.

\section{Determination of expansion coefficients}
\subsection{Initialization}
In this section we explain how Mathematica can be used to evaluate
all the expansion coefficients needed for further implementation of
matrix models. The method is explained with great generality and can
be used up to any degree of approximation and with all surfaces that
can be defined analytically.

First of all we must load two packages that will be used further
along:
\begin{verbatim}
 Needs["Calculus`VectorAnalysis`"]
 Needs["Optics`expansion`"]
\end{verbatim}
The first package is distributed standard with Mathematica and is
useful for performing vector operations. The second one is a
proprietary package which performs series expansion in different
way. The standard series expansion for multi-variable functions sets
the limit for each variable's exponent independently but not for the
order of the monomials. The \emph{expansion} package sets a limit
for the monomials' order, so that each variable's exponent is
adjusted accordingly; for details see appendix \ref{s:expackage}.

\subsection{Surface definition}
The method is applicable to any surface which can be defined by an
equation of the type $\mathrm{f}(x,y,z)=0$. For instance, a conic
of revolution could be defined as:
\begin{verbatim}
 surface[x_, y_, z_] = Sqrt[x^2 + y^2 + z^2] + e*z - r;
\end{verbatim}

We are going to need the normal to the surface at any point; this
can be found applying gradient to the surface expression:
\begin{verbatim}
 n = Grad[surface[Xx, Yy, Zz], Cartesian[Xx,Yy,Zz]];
\end{verbatim}

\subsection{\label{s:Snel}Snell's law}
Snell's law is usually written as scalar relationship between sines
of the angles the rays form with the normal on both sides of the
surface, but it can also be expressed as a vector relationship using
cross products \cite{Almeida99}. Let us define to unit vectors
representing the ray directions on the two sides of the surface:
\begin{verbatim}
 v = {s, t, Sqrt[1 - s^2 - t^2]};
 v1 = {s1, t1, Sqrt[1-s1^2-t1^2]};
\end{verbatim}
\texttt{v} and \texttt{v1} represent the unit vectors, \texttt{s},
\texttt{t}, \texttt{s1} and \texttt{t1} are direction cosines.

Snell's law requires that the cross product of the ray direction
with the surface normal, multiplied by the refractive index be equal
on both sides:
\begin{equation}\label{eq:Snel}
  u \,{\bf v} \otimes {\bf n} = {\bf v1} \otimes {\bf n}~,
\end{equation}
where $u$ represents the refractive index ratio of the two media.
We ask Mathematica to find the solutions of Eq.\ (\ref{eq:Snel})
for the surface in question:
\begin{verbatim}
 Snell = Solve[{u*CrossProduct[v, n] ==
       CrossProduct[v1, n],surface[Xx,Yy,Zz]==0}, {s1, t1,Zz}];
\end{verbatim}

Eq.\ (\ref{eq:Snel}) has four solutions, corresponding to positive
and negative angles, in the forward and the reverse directions.
The solutions found by Mathematica must be scanned in order to
find the appropriate one. We can write the exact solution with two
commands:
\begin{verbatim}
 sd = Simplify[s1 /. Snell[[3]]]
 td = Simplify[t1 /. Snell[[3]]]
\end{verbatim}
which will output very long expressions very impractical for use.
\subsubsection{\label{s:expansion}Series expansion} The solutions of Eq.\
(\ref{eq:Snel}) must be expanded in series to the desired order of
approximation for which we use a function from the
\emph{expansion} package. Dealing with axis-symmetric systems it
is useful to resort to complex coordinates; the two solutions are
thus combined into one single complex solution, then expanded to
the fifth order and simplified:
\begin{verbatim}
 sexpand = Expansion[sd + I*td, s, t, Xx, Yy, 5];
 sexpand = Simplify[PowerExpand[sexpand]];
\end{verbatim}

The result of the \texttt{Expansion} function is a list of
coefficients, most of them zero. The \emph{expansion} package
provides a function \texttt{VectorBase} which outputs a list of
all the monomials that can be built with the given list of
variables up to the desired order. This can be left-multiplied by
the coefficient list in order to get a polynomial:
\begin{verbatim}
 sexpand = sexpand . VectorBase[s, t, x, y, 5];
\end{verbatim}

\subsubsection{\label{s:complex}Complex coordinates}
Although combined into one single complex coordinate in
\texttt{sexpand}, the output ray direction cosines are still
expressed in terms of real input ray position and direction
coordinates; this must be corrected by an appropriate coordinate
change, followed by suitable manipulation and simplification:
\begin{verbatim}
 Apply[Plus,Simplify[MonomialList[ComplexExpand[sexpand /.
   {x -> (X + Conjugate[X])/2,
    y -> (X - Conjugate[X])/(2*I),
    s -> (S + Conjugate[S])/2,
    t -> (S - Conjugate[S])/(2*I)}, {X,S},
    TargetFunctions->Conjugate],
    {X, Conjugate[X], S, Conjugate[S]},
    CoefficientDomain->RationalFunctions]]]
\end{verbatim}
Mathematica will output an expression for the above command, which
is reproduced bellow in a slightly different form:
\begin{eqnarray*}
&&S\,u - {\frac{\left( 1 + u \right) \,X}
    {r}} + {\frac{S\,u\,
      \left( 1 + u \right) \,X\,
      \mathrm{Conjugate}(S)}{2\,r}} -
  {\frac{u\,\left( 1 + u \right) \,{X^2}\,
      \mathrm{Conjugate}(S)}{2\,{r^2}}}\\
   &&+ {\frac{{S^2}\,
      \left( u + {u^4} \right) \,X\,
      {{\mathrm{Conjugate}(S)}^2}}{8\,r}
    } - {\frac{S\,{u^2}\,
      \left( -1 + {u^2} \right) \,{X^2}\,
      {{\mathrm{Conjugate}(S)}^2}}{4\,
      {r^2}}}\\
      &&+ {\frac{{u^2}\,
      \left( -1 + {u^2} \right) \,{X^3}\,
      {{\mathrm{Conjugate}(S)}^2}}{8\,
      {r^3}}} - {\frac{S\,u\,
      \left( 1 + u \right) \,X\,
      \mathrm{Conjugate}(X)}{2\,{r^2}}}\\
   &&+ {\frac{\left( 1 + u \right) \,
      \left( {e^2} + u \right) \,{X^2}\,
      \mathrm{Conjugate}(X)}{2\,{r^3}}}\\
    &&- {\frac{{S^2}\,{u^2}\,
      \left( -1 + {u^2} \right) \,X\,
      \mathrm{Conjugate}(S)\,
      \mathrm{Conjugate}(X)}{4\,{r^2}}}\\
    &&- {\frac{S\,u\,\left( 1 + u \right) \,
      \left( 1 + {e^2} + 2\,u -
        2\,{u^2} \right) \,{X^2}\,
      \mathrm{Conjugate}(S)\,
      \mathrm{Conjugate}(X)}{4\,{r^3}}}\\
   &&+ {\frac{u\,\left( 1 + u \right) \,
      \left( 2\,{e^2} + u - {u^2} \right)
        \,{X^3}\,\mathrm{Conjugate}(S)\,
      \mathrm{Conjugate}(X)}{4\,{r^4}}}\\
   &&+ {\frac{{S^2}\,{u^2}\,
      \left( -1 + {u^2} \right) \,X\,
      {{\mathrm{Conjugate}(X)}^2}}{8\,
      {r^3}}}\\
      &&+ {\frac{S\,u\,
      \left( 1 + u \right) \,
      \left( 2\,{e^2} + u - {u^2} \right)
        \,{X^2}\,{{\mathrm{Conjugate}(
          X)}^2}}{4\,{r^4}}}\\
          &&+  {\frac{\left( 1 + u \right) \,
      \left( -3\,{e^4} + u - 6\,{e^2}\,u -
        {u^2} + {u^3} \right) \,{X^3}\,
      {{\mathrm{Conjugate}(X)}^2}}{8\,
      {r^5}}}
\end{eqnarray*}

All the coefficients resulting from similar operations performed
on a spherical surface up to the seventh order have already been
listed by the author \cite{Almeida99} and the reader is referred
to that work for further elucidation.

\subsection{Surface offset}
We will now deal with the problem of the offset introduced by the
surface on the position coordinates; the reader is again referred
to the above mentioned work for  full explanation of this matter.

Mathematica will respond quicker if we perform shift the coordinate
origin, along the axis, to the geometrical centre of the surface;
this is the purpose of the three following commands:
\begin{verbatim}
 eq = Solve[surface[x, y, z] == 0, z],
 z1 = z /. eq[[2]];
 z1 = z1 - r/(-1 + e);
\end{verbatim}

The ray coordinates on the point of incidence will have to obey
the two equations:
\begin{eqnarray}
\label{eq:offset1}
    x1&=& x + {\frac{s z1}{ \sqrt{1 - s^2 -
     t^2}}}~;\nonumber\\
    y1&=& y + {\frac{t z1}{ \sqrt{1 - s^2 -
     t^2}}}~.
\end{eqnarray}
The following commands will perform the necessary calculations:
\begin{verbatim}
 eqx = x1 == x + s*z1/Sqrt[1 - s^2 - t^2];
 eqy = y1 == y + t*z1/Sqrt[1 - s^2 - t^2];
 offset = Solve[{eqx, eqy}, {x, y}];
\end{verbatim}

The ray will usually intersect a surface in two points but complex
surfaces can be intersected in several points; the relevant
solution is the one that is closest to the surface vertex plane,
which will have to be selected from the multiple solutions found
by Mathematica:
\begin{verbatim}
 xd = Simplify[x /. offset[[2]]];
 yd = Simplify[y /. offset[[2]]];
\end{verbatim}

\subsubsection{Series expansion}
The commands that follow reproduce a procedure similar to what was
explained in paragraphs \ref{s:expansion} and \ref{s:complex}:
\begin{verbatim}
 xexpand =
  Simplify[PowerExpand[Expansion[xd + I*yd, s,
     t, x1, y1, 5]],TimeConstraint->Infinity];
 xexpand = xexpand . VectorBase[s, t, x, y, 5];
 Simplify[MonomialList[ComplexExpand[xexpand /.
   {x -> (X + Conjugate[X])/2,
    y -> (X - Conjugate[X])/(2*I),
    s -> (S + Conjugate[S])/2,
    t -> (S - Conjugate[S])/(2*I)}, {X,S},
    TargetFunctions->Conjugate],
    {X, Conjugate[X], S, Conjugate[S]},
    CoefficientDomain->RationalFunctions]]
\end{verbatim}
The result is the following coefficient list:
\begin{eqnarray*}
&\{& {\frac{\left( -1 + {e^2} \right) \,S\,
      {X^2}\,{{\mathrm{Conjugate}(X)}^2}
      }{8\,{r^3}}}, \\
  &&{\frac{S\,{X^2}\,
      \mathrm{Conjugate}(S)\,
      \mathrm{Conjugate}(X)}{4\,{r^2}}}, \\
  &&{\frac{{S^2}\,X\,
      {{\mathrm{Conjugate}(X)}^2}}{4\,
      {r^2}}}, \\
      &&{\frac{-\left( {S^2}\,X\,
        \mathrm{Conjugate}(S)\,
        \mathrm{Conjugate}(X) \right) }
      {4\,r}}, \\
      &&{\frac{-\left( S\,X\,
        \mathrm{Conjugate}(X) \right) }
      {2\,r}},X\, \}
\end{eqnarray*}

The complete set of coefficients for the seventh-order and
spherical surfaces can be found on the paper mentioned before. The
same reference lists the coefficients for the reverse offset,
which are determined in an entirely similar manner.

\subsection{Straight path} The use of direction cosines to define
the ray orientation renders a straight path into a non-linear
transformation, whose expansion must also be taken care of. The
procedure is similar to what was used above and there are no
mathematical complexities involved; we will just list the commands
and the final result:
\begin{verbatim}
 xexpand =
  Simplify[PowerExpand[Expansion[x +
      s*e/Sqrt[1 - s^2 - t^2] +
      I*(y + t*e/Sqrt[1 - s^2 - t^2]), s, t, x,
     y, 5]]];
 xexpand = xexpand . VectorBase[s, t, x, y, 5];
 Apart[xexpand /.
   {x -> (X + Conjugate[X])/2,
    y -> (X - Conjugate[X])/(2*I),
    s -> (S + Conjugate[S])/2,
    t -> (S - Conjugate[S])/(2*I)}, r]
\end{verbatim}
The final output is:
\begin{eqnarray*}
{\frac{1}{8}\left({8\,e\,S + 8\,X +
     4\,e\,{S^2}\,
      \mathrm{Conjugate}(S) +
     3\,e\,{S^3}\,
      {{\mathrm{Conjugate}(S)}^2}}\right)}
\end{eqnarray*}

\section{Simulation of a Petzval lens}

\begin{figure}[htb]
    \centerline{\psfig{file=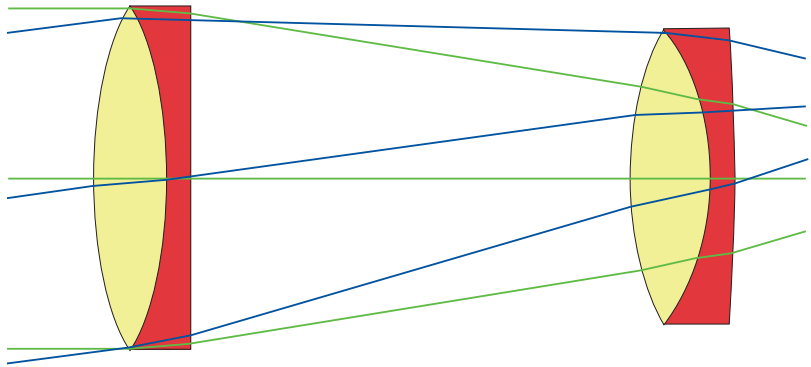, scale=0.7}}
\caption{\label{fig:petzval} Diagram of a Petzval lens.}
\end{figure}

In this section we will use Mathematica to model a complex lens
using matrix optics. The chosen lens is a Petzval design
distributed with Oslo LT \cite{Oslo95}, which reproduces an
example from Walker \cite{Walker95}; this lens is illustrated in
Fig.\ \ref{fig:petzval} and is built with four elements made of
glasses BK7 and SF4.
\subsection{Initialization}
The model uses a proprietary package named \emph{Sphericl} which
defines all the functions needed for implementation of matrix
models of spherical systems, along with some other useful
functions. The package is described in appendix \ref{s:sphericl}.
\begin{verbatim}
 Needs["Optics`Sphericl`"]
\end{verbatim}

The two glasses used for the individual elements are defined by
lists of their refractive indices at three wavelengths, namely
{587.56~nm}, {486.130~nm} and {656.270~nm}:
\begin{verbatim}
 bk7 = {1.5168, 1.522376, 1.514322};
 f4 = {1.616592, 1.62848, 1.611645};
\end{verbatim}

\subsection{Lens definition}
The two functions \texttt{Lens[]} and \texttt{Distance[]} provide
the matrices corresponding to the ray transformations induced by a
single element in air and a straight path; they are multiplied
successively, in reverse order relative to the transformations, in
order to model the lens:
\begin{verbatim}
 Table[lens[i] =
    Lens[f4[[i]],-18.44,-40.23,3] . Distance[35] .
    Lens[bk7[[i]],41.42,103.08,8] . Distance[25] .
    Lens[bk7[[i]],-72.05,-92.6,7] . Distance[4] .
    Lens[bk7[[i]],53.12,-799.9,10] . Distance[60],{i,1,3}];
\end{verbatim}
The last factor, \texttt{Distance[60]}, corresponds to the
distance from the aperture stop to the first surface. The
\texttt{Table[]} function is used to generate models for all three
wavelengths.

\subsection{Image coordinates}
The optimization performed by Oslo determined an image plane
position at {3.787202~mm} from the last surface, so there is one
last product to make in order to find the overall system matrices
for the three wavelengths:
\begin{verbatim}
 Table[image[i] = Distance[3.787202].lens[i],{i,1,3}];
\end{verbatim}

The system matrices must be right-multiplied by the vector base in
order to generate the coordinates and the higher order monomials
for the image plane. The function \texttt{Terms[]} is responsible
for generating a complex vector base from the given coordinates:
\begin{verbatim}
 Table[imagex[i]=(image[i].Terms[{x,s}])[[1]],{i,1,3}];
\end{verbatim}
List \texttt{imagex[i]} is a 3 element vector, whose elements are
the complex position coordinates for each of the three wavelengths.
We can isolate the tangential and sagittal components by making the
input coordinate real or imaginary and simultaneously taking the
real or imaginary parts, respectively:
\begin{verbatim}
 Table[timage[i]=ComplexExpand[Re[imagex[i]]],{i,1,3}];
 Table[simage[i]=ComplexExpand[Im[imagex[i]/. x->I x]],{i,1,3}];
\end{verbatim}

\subsection{Ray analysis}
\begin{figure}[htb]
    \centerline{\psfig{file=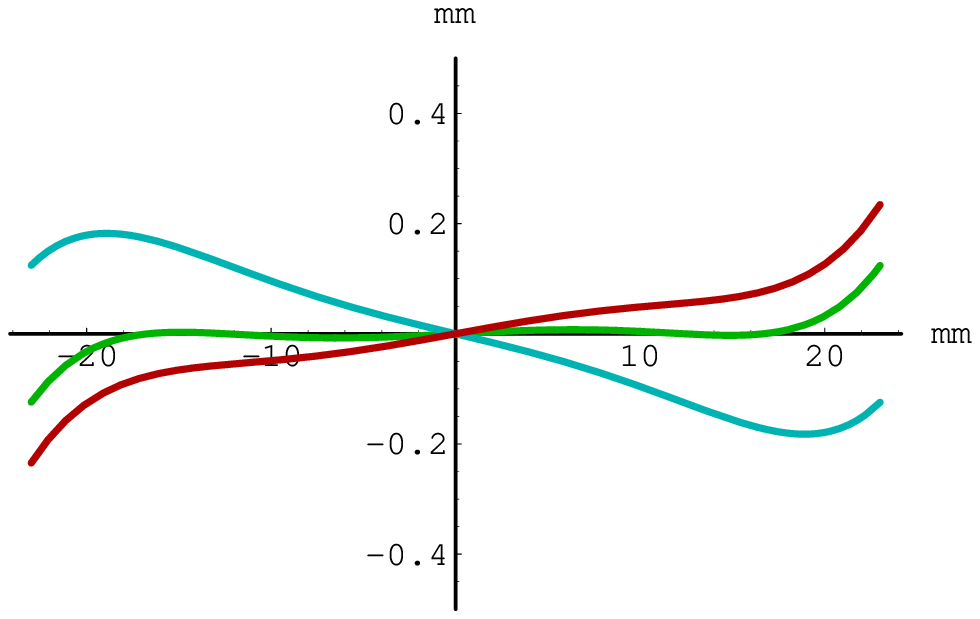, scale=0.7}\hspace{1cm}
    \psfig{file=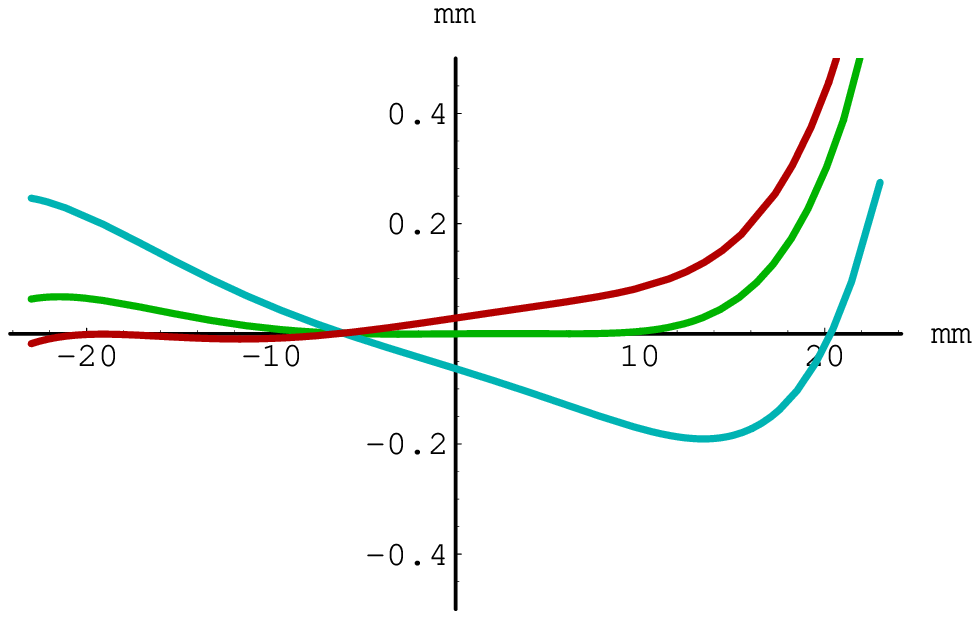, scale=0.7}}
\caption{\label{fig:intercept} Ray intercept curves for three
wavelengths: green line -- {587.56~nm}, blue line -- {486.130~nm},
red line -- {656.270~nm}. Left figure is for on-axis incidence,
right figure for $5^\mathtt{o}$ field angle.}
\end{figure}

Finally we perform some ray analysis by plotting ray intercept
curves for two different values of the field given by the variable
\texttt{s}; the results are shown in Fig.\ \ref{fig:intercept}:
\begin{verbatim}
 Table[timage0 = timage[1] /. x -> 0;
  Plot[{timage[1] - timage0, timage[2] - timage0,
      timage[3] - timage0}, {x, -23, 23},
    PlotRange -> {Automatic, {-0.5, 0.5}},
    PlotStyle -> {{AbsoluteThickness[2],
          RGBColor[0, 0.7, 0]}, {AbsoluteThickness[2],
          RGBColor[0, 0.7, 0.7]}, {AbsoluteThickness[2],
          RGBColor[0.7, 0, 0]}}, AxesStyle -> AbsoluteThickness[1],
    AxesLabel -> {"mm", "mm"}], {s, 0, 0.09, 0.087155}]
\end{verbatim}

\section{Conclusion}
All the calculations needed for matrix modelling optical systems
were successively programmed with Mathematica without limitations
for surface shapes or degree of approximation. The program for the
determination of series expansion coefficients was run in less than
one hour for the seventh-order and spherical surfaces. Other surface
shapes have already bean used and these include conicals and
toroids.

After calculation, the coefficients have been incorporated in a
Mathematica package which runs fast and avoids the need to
re-calculate over and over. This package is fast and allows the
simulation of very complex optical systems; one such example was
demonstrated in the form of a Petzval lens.

Work is now going on to extract more possibilities from the
software, namely for plotting wavefronts and ray-densities, and
will be the object of other publications.
\appendix
\section{\label{s:sphericl}The ''\texttt{Sphericl}'' package}
This package provides all the functions needed for the
implementation of seventh-order matrix models of optical systems.
Some auxiliary functions are also defined in order to facilitate
other optical system calculations. This appendix describes in
detail the fundamental functions and gives only short mentions of
the others.

The package makes use of the axis symmetry to reduce matrix size
to $40 \times 40$, so all the coordinates are assumed to be
complex. One ray, at any specific position is characterized by one
complex position coordinate and on complex direction cosine
coordinate.

\subsection*{\texttt{Terms[]}}
The function builds the 40-element vector base with all the
coordinate monomials that have non-zero coefficients in
axis-symmetric systems:

\begin{verbatim}
 Terms[r_List] := Module[{x,s,ray},
 s = r[[2]] ; x = r[[1]]; ray = {x, s, x^2*Conjugate[x],
 x^2*Conjugate[s], s*x*Conjugate[x],
  s*x*Conjugate[s], s^2*Conjugate[x], s^2*Conjugate[s],
  x^3*Conjugate[x]^2, x^3*Conjugate[s]*Conjugate[x],
  x^3*Conjugate[s]^2, s*x^2*Conjugate[x]^2,
  s*x^2*Conjugate[s]*Conjugate[x], s*x^2*Conjugate[s]^2,
  s^2*x*Conjugate[x]^2, s^2*x*Conjugate[s]*Conjugate[x],
  s^2*x*Conjugate[s]^2, s^3*Conjugate[x]^2,
  s^3*Conjugate[s]*Conjugate[x], s^3*Conjugate[s]^2,
  x^4*Conjugate[x]^3, x^4*Conjugate[s]*Conjugate[x]^2,
  x^4*Conjugate[s]^2*Conjugate[x], x^4*Conjugate[s]^3,
  s*x^3*Conjugate[x]^3, s*x^3*Conjugate[s]*Conjugate[x]^2,
  s*x^3*Conjugate[s]^2*Conjugate[x], s*x^3*Conjugate[s]^3,
  s^2*x^2*Conjugate[x]^3,
  s^2*x^2*Conjugate[s]*Conjugate[x]^2,
  s^2*x^2*Conjugate[s]^2*Conjugate[x],
  s^2*x^2*Conjugate[s]^3, s^3*x*Conjugate[x]^3,
  s^3*x*Conjugate[s]*Conjugate[x]^2,
  s^3*x*Conjugate[s]^2*Conjugate[x], s^3*x*Conjugate[s]^3,
  s^4*Conjugate[x]^3, s^4*Conjugate[s]*Conjugate[x]^2,
  s^4*Conjugate[s]^2*Conjugate[x], s^4*Conjugate[s]^3};
 ray];
\end{verbatim}

\subsection*{\texttt{otherlines[]}}
All the transformation matrices have a common structure
characterized by two lines of series expansion coefficients
followed by thirty eight lines of elements derived from the former
by algebraic operations. The \texttt{otherlines[]} function is an
internal function which accepts as input any $40 \times 40$ matrix
and outputs another matrix with the same first two lines and lines
3 to 40 built according to the common rules.

It is impossible to list all the implementation lines for this
function, for reasons of size; we will then resort to an
explanation of the procedures for determining the coefficients for
one of the 38 lines.

Suppose we want to find the coefficients for a line which
corresponds to the monomial $X^j \mathrm{Conjugate}\left[X
\right]^k S^l \mathrm{Conjugate}\left[S \right]^m$. We start be
defining a square matrix \texttt{t} of dimension 40, with known
elements on the first two rows:
\begin{verbatim}
 t = IdentityMatrix[40];
 t[[1]]=Table[a[i],{i,1,40}];
 t[[2]]=Table[b[i],{i,1,40}];
\end{verbatim}

The matrix \texttt is right-multiplied by the vector base of
coordinate monomials and the first two elements of the product are
isolated:
\begin{verbatim}
 X1 = (t . Terms[{X, S}])[[1]];
 S1 = (t . Terms[{X, S}])[[2]];
\end{verbatim}
The procedure then involves determining the product
\begin{verbatim}
 X1^j Conjugate[X1]^k S1^l Conjugate[S1]^m
\end{verbatim}
and selecting just the terms up to the seventh order. The same
procedure must be repeated for all the 38 lines.

These are lengthy calculations which must be performed only once.
The function \texttt{otherlines[]} incorporates the results of
those calculations and is fast to operate.

\subsection*{\texttt{refraction[]}}
This is an internal function which receives as input the refractive
indices of the two media and the curvature radius and outputs a
transformation matrix whose second line contains the expansion
coefficients for Snell's law according to paragraph \ref{s:Snel}:

\begin{verbatim}
 refraction[n1_, n2_, r_] := Module[{matrix,u},
 matrix = IdentityMatrix[40];
 u=n1/n2;
 matrix[[2]] =
 {(u-1)/r,u,(u^2-u)/(2 r^3),(u^2-u)/(2 r^2),(u^2-u)/(2
 r^2),(u^2-u)/(2 r),0,0, (u^4-u)/(8 r^5),(u^4-u^2)/(4
 r^4),(u^4-u^2)/(8 r^3),(u^4-u^2)/(4 r^4),u(2u^3-3 u + 1)/(4 r^3),
 (u^4 - u^2)/(4 r^2),(u^4-u^2)/(8 r^3),(u^4-u^2)/(4 r^2),(u^4-u)/(8
 r),0,0,0, (u^6-u)/(16 r^7),u^2 (3u^4 - 2u^2 - 1)/(16 r^6),3
 u^4(u^2-1)/(16 r^5),(u^6-u^4)/(16 r^4), u^2(3 u^4-2 u^2-1)/(16
 r^6),u(9u^5-10u^3 + 1)/(16 r^5),u^2(9 u^4- 11 u^2 + 2)/(16 r^4), 3
 u^4(u^2-1)/(16 r^3),3 u^4(u^2-1)/(16 r^5),u^2(9 u^4-11 u^2+2)/(16
 r^4), u(9 u^5-10 u^3+1)/(16 r^3),u^2(3 u^4-2 u^2-1)/(16
 r^2),u^4(u^2-1)/(16 r^4),3 u^4(u^2-1)/(16 r^3), u^2(3 u^4-2
 u^2-1)/(16 r^2),u(u^5-1)/(16 r),0,0,0,0};
 otherlines[matrix]];
\end{verbatim}

\subsection*{\texttt{Screen[]}}
This is an external function which is used internally to determine
the surface offset and externally to deal with spherical image
surfaces. The structure is similar to the previous one but now its
the first line which is defined:

\begin{verbatim}
 Screen[r_] := Module[{matrix},
 matrix = IdentityMatrix[40];
 matrix[[1]] = {1,0,0,0,1/(2r),0,0,0,0,0,0,
 1/(8 r^3),0,0,0,1/(4 r),0,0,0,0,0,0,0,0,1/(16
 r^5),0,0,0,0,1/(16 r^3),0,0,0,0,3/(16 r),0,0,0,0,0};
 otherlines[matrix]];
\end{verbatim}

\subsection*{\texttt{back[]}}
This is an internal function used to determine the reverse offset.

\begin{verbatim}
 back[r_] := Module[{matrix},
 matrix = IdentityMatrix[40];
 matrix[[1]] = {1,0,0,0,-1/(2r),0,0,0,0,0,0,
 -1/(8 r^3),1/(4 r^2),0,1/(4 r^2),-1/(4r),
 0,0,0,0,0,0,0,0,-1/(16 r^5),3/(16 r^4),
 -1/(8 r^3),0,3/(16r^4),-7/(16 r^3),1/(4 r^2),
 0,-1/(8 r^3),1/(4 r^2),-3/(16r),0,0,0,0,0};
 otherlines[matrix]];
\end{verbatim}

\subsection*{\texttt{Surface[]}}
This is an external function which receives as input the refractive
indices of the two media and outputs the transformation matrix for a
surface; it is built as the product of three matrices:

\begin{verbatim}
 Surface[n1_, n2_, r_] :=
 back[r] . refraction[n1, n2, r] . Screen[r];
\end{verbatim}

\subsection*{\texttt{Distance[]}}
This is an external function which receives as input the distance
travelled along the axis and outputs the matrix for the straight
path transformation:

\begin{verbatim}
 Distance[t_] := Module[{matrix},
 matrix = IdentityMatrix[40];
 matrix[[1]] = {1,t,0,0,0,0,0,t/2,0,0,0,0,0,
 0,0,0,0,0,0,3t/8,0,0,0,0,0,0,0,0,0,0,0,0,0,
 0,0,0,0,0,0,5 t/16};
 otherlines[matrix]];
\end{verbatim}

\subsection*{Other functions}
Auxiliary functions include \texttt{Lens[]}, for the matrix of a
single element lens in air, functions for gaussian constants,
conjugates, etc.
\section{\label{s:expackage}The ''\texttt{expansion}'' package}
\begin{verbatim}
BeginPackage["Optics`expansion`"]

Expansion::usage =
    "Expansion[e,{x,y,z,t},o] finds the coefficients of the series
    expansion of expression e in variables x,y,z,t up to the order o";


Begin["`Private`"];

Expansion[a_, {ax_, ay_, rx_, ry_}, o_] :=
    Module[{final, ser1, ser2, ser3, ser4}, final = 0;
      ser1 = Series[a, {ax, 0, o}];
      For[i = 0, i < o + 1, i++,
        ser2 = Series[Coefficient[ser1, ax, i], {ay, 0, o}];
        For[j = 0, j < o + 1 - i, j++,
          ser3 = Series[Coefficient[ser2, ay, j], {rx, 0, o}];
          For[k = 0, k < o + 1 - i - j, k++,
            ser4 = Series[Coefficient[ser3, rx, k], {ry, 0, o}];
            For[u = 0, u < o + 1 - i - j - k, u++,
              final=final+Coefficient[ser4,ry,u]*ax^i*ay^j*rx^k*ry^u;
              ]]]]; final]



End[]; EndPackage[];
\end{verbatim}
%
%\pagebreak
  \bibliography{aberrations}   %>>>> bibliography data in aberrations.bib
  \bibliographystyle{spiebib}   %>>>> makes bibtex use spiebib.bst

\end{document}